\documentclass[12pt]{article}
\usepackage{amsmath}
\usepackage{graphicx,psfrag,epsf}
\usepackage{enumerate}
\usepackage{natbib}
\usepackage{hyperref}
\usepackage{url} 
\usepackage[table]{xcolor}
\usepackage{rotating}
\usepackage{tabularx}
\usepackage{ragged2e}
\usepackage{multirow}
\usepackage{caption}

\newcommand{\blind}{0}

\newcommand{\bftab}{\fontseries{b}\selectfont}

\begin{document}

\def\spacingset#1{\renewcommand{\baselinestretch}%
{#1}\small\normalsize} \spacingset{1}


\if0\blind
{
  \title{\bf ``6 choose 4'': A framework to understand and facilitate discussion of strategies for overall survival safety monitoring}
  \author{Godwin Yung\\
    F. Hoffmann-La Roche AG, South San Francisco, CA, US\\
    and \\
    Kaspar Rufibach \\
    F. Hoffmann-La Roche AG, Basel, Switzerland\\
    and \\
    Marcel Wolbers \\
    F. Hoffmann-La Roche AG, Basel Switzerland \\
    and \\
    Mark Yan \\
    F. Hoffmann-La Roche AG, Mississauga, CAN \\
    and \\
    Jue Wang \\
    F. Hoffmann-La Roche AG, South San Francisco, CA, US}
  \maketitle
} \fi

\if1\blind
{
  \bigskip
  \bigskip
  \bigskip
  \begin{center}
    {\LARGE\bf ``6 choose 4'': A framework to understand and facilitate discussion of strategies for overall survival safety monitoring}
\end{center}
  \medskip
} \fi

\bigskip
\begin{abstract}
Advances in anticancer therapies have contributed significantly to decreasing death rates in certain disease and clinical settings. However, they have also made it difficult to power a clinical trial in these settings with overall survival (OS) as the primary efficacy endpoint. Therefore, two approaches have recently been proposed for the pre-specified analysis of OS as a safety endpoint \citep{fleming, rodriguez}. We provide in this paper a simple unifying framework that includes the aforementioned approaches---and a few others---as special cases. By highlighting each approach's focus, priority, tolerance for risk, and strengths or challenges for practical implementation, this framework can help facilitate discussions between stakeholders on ``fit-for-purpose OS data collection and assessment of harm'' \citep{aacr}. We apply this framework to a real clinical trial in large B-cell lymphoma to illustrate its application and value. Several recommendations and open questions are also raised.
\end{abstract}

\noindent%
\vfill

\newpage
\spacingset{1.45} 
\section{Introduction}
\label{sec:intro}

With the continued advancement of anticancer therapies, there are now many disease and clinical contexts in which patients' lives are extended and multiple treatment options are available following disease progression. Although these successes should be celebrated, they have also made it challenging for clinical trials to have adequate power to detect statistically significant improvements in overall survival (OS)---the gold standard endpoint in oncology \citep{Merino2024}. In July 2023, a workshop titled ``Overall Survival in Oncology Clinical Trials'' was held jointly by the U.S. Food and Drug Administration (FDA), American Association for Cancer Research (AACR), and American Statistical Association (ASA) to discuss challenges with the timely evaluation of OS for novel therapies \citep{aacr}. The workshop cited lack of OS data, lack of plans for further OS data collection, and lack of pre-specified OS analyses as emerging challenges for regulatory authorities to interpret OS data in trials where OS is not a primary or key secondary endpoint. A call to action was made to ``encourage thoughtful and comprehensive planning during trial design for fit-for-purpose OS data collection and assessment of harm given the disease and clinical context''.

Following that call-to-action, two statistical approaches were proposed for the pre-specified analysis of OS as a safety endpoint. \cite{fleming} proposed a monitoring guideline inspired by ones used by the FDA in cardiovascular safety trials. \cite{rodriguez} suggested using confidence intervals to decide whether or not there is sufficient evidence to rule out harm. 
These approaches are similar to a certain degree. For example, both consider similar sets of parameters: definition of harm measured in terms of the OS hazard ratio, type I error, etc.. However, upon closer look, it can be unclear even to the experienced statistician how these approaches relate to each other exactly, how they differ, which approach one should use in practice, and why.

This paper proposes a framework for OS safety analysis that includes as special cases \cite{fleming} and \cite{rodriguez}. Importantly, this framework explains what flexibility and restrictions exist for \emph{any} approach analyzing OS as a safety endpoint under the traditional framework of hypothesis testing, and clarifies for each approach their focus, priority, and tolerance for risk over the course of a trial. By increasing understanding of different approaches and the relationship between them, this framework can help facilitate ``thoughtful planning'' by sponsors and improve communication between stakeholders in order to achieve alignment on clinical trial design and analysis.

The rest of the paper is organized as follows. In Section \ref{sec:meth}, we present our framework for OS safety monitoring and provide five examples of approaches that fall under this framework, including \cite{fleming} and \cite{rodriguez}. In Section \ref{sec:eg}, we illustrate and compare the five approaches by applying them to POLARIX, a real clinical trial studying patients with untreated large B-cell lymphoma. Section \ref{sec:discussion} concludes with recommendations for practice and open questions for further deliberation or research.

\section{Proposed framework}
\label{sec:meth}

Consider a trial with $n$ patients, $n\pi$ randomized to the experimental arm and $n(1-\pi)$ randomized to the control arm. A prospectively designed hypothesis test for OS detriment---as measured by the OS hazard ratio (HR)---involves six parameters: 
\begin{enumerate}
\item $\theta_0$ the OS HR under the null hypothesis $H_0$ of OS detriment
\item $\theta_1$ the OS HR under the alternative hypothesis $H_1$ of no OS detriment
\item $d$ the number of deaths
\item $\theta^*$ the threshold for decision making between $H_0$ vs.\ $H_1$
\item $\alpha$ the one-sided type I error rate
\item $\beta$ the type II error rate
\end{enumerate}
For the purpose of safety monitoring in indolent cancer trials, we may limit ourselves to considering $\theta_0 > 1.0$ and moderate to neutral $\theta_1$ (e.g., between 0.7 and 1.0). \cite{fleming} characterized $\theta_0$ as ``the smallest unacceptable detrimental OS HR'' and $\theta_1$ as ``a plausible OS HR consistent with benefit that is reasonably expected from the experimental intervention''.

Two classical equations characterize the approximate relationship between the six parameters:
\begin{equation}
\alpha = Pr(\hat \theta < \theta^*|\theta_0) = \Phi \left ( \log(\theta^*/\theta_0) \sqrt{\pi(1-\pi)d} \right )
\label{eq:t1e}
\end{equation}
\begin{equation}
1-\beta = Pr(\hat \theta < \theta^*|\theta_1) = \Phi \left ( \log(\theta^*/\theta_1) \sqrt{\pi(1-\pi)d} \right )
\label{eq:power}
\end{equation}
where $\Phi(\cdot)$ denotes the cumulative distribution function of the standard normal distribution. These equations assume proportional hazards. More specifically, they assume that the observed log hazard ratio $\log(\hat \theta)$ follows a normal distribution with variance $1/(\pi(1-\pi)d)$ and mean $\log(\theta_0)$ or $\log(\theta_1)$ \citep{schoenfeld}.

Given six parameters and Equations (\ref{eq:t1e})-(\ref{eq:power}) characterizing the relationship between them, our proposed framework is as follows: \emph{at each analysis time, users input or ``choose'' 4 of the 6 parameters (not all from the same equation) and solve for the remaining 2.} This one-sentence framework communicates clearly what flexibility and restrictions exist in any OS monitoring strategy. It highlights a strategy's focus and associated risks; for example, choosing $\beta$ and solving for $\alpha$ at an early interim analysis implies a focus on having the trial continue in case the experimental therapy is effective, while understanding the risk of a false positive. And as we shall illustrate, it includes as special cases \cite{fleming}, \cite{rodriguez}, and other approaches for OS safety monitoring. 

Before proceeding, there are three important points that deserve clarification. First, $\alpha$ and $\beta$ here refer to error rates associated with a single test as opposed to a group-sequential test and/or multiple endpoints. More specifically, they refer to \emph{marginal} type I and type II error rates for hypothesis testing of OS \emph{safety} at interim and final analyses. They should not be confused with \emph{overall} error rates across all analysis times. (Note that Fleming et al.\ refer to $\alpha$ and $\beta$ as false positive and false negative error rates. We prefer the terms marginal type I and type II error rates to more clearly distinguish them from overall error rates.) They should also not be confused with error rates for primary and secondary endpoints with a formal testing plan. Whether OS should be formally tested as a safety endpoint and how to do so in relation to other formally tested endpoints is context specific and deserves careful consideration in practice. However, a thorough discussion of this topic is beyond the scope of this paper.

Second, the proposed framework should not be misunderstood as suggesting a single one-time calculation. Just because a user inputs certain parameters does not mean they should accept whatever the other parameters end up being. All six parameters are important and an iterative process is likely needed to arrive at a configuration that various stakeholders can agree on. 
The intention of our framework is not to encourage focus on some parameters while ignoring others, but rather to elucidate and encourage stakeholder discussion of the relative priorities and trade-offs between the parameters. Our framework may also help clarify how certain parameters are more informed by data or clinical/scientific considerations compared to other parameters.

Third, the word ``choose'' should not be misunderstood as suggesting that sponsors are always at liberty to specify the value of a parameter. Similar to the previous note, certain parameter values may be driven by trial conduct, clinical consideration, or regulatory context. For example, in a trial of indolent cancer where timing of the first analysis is based on progression free survival (PFS), the number of OS events at this time cannot be prespecified. It will be what-it-is when the prespecified number of PFS events is observed. At the time of study design, we can input a \emph{predicted} number of OS events to explore operating characteristics of a safety monitoring strategy. However, when the actual trial takes place, the input will need to be updated based on the \emph{observed} number of OS events.  If the observed number of OS events is quite different from the predicted, then the entire strategy and configuration may need to be revisited to ensure that stakeholders are realigned on the focus and risks.

\subsection*{Special case \#1: Fleming et al. (2024)}
Fleming et al.'s OS monitoring guideline can be understood under the proposed framework as choosing $(\theta_0, \theta_1, \beta, d)$ at interim analyses (IAs) and $(\theta_0, \theta_1, \alpha, d)$ at the final analysis (FA). Their approach places more emphasis on trial continuation while evidence is still accumulating, which could make sense in trials with limited OS events at early IAs. This is later counter-balanced by a focus on ruling out OS detriment with reasonably low type I error rate $\alpha$ when more OS events have accumulated. 

While representing Fleming et al.’s guideline in such a way makes it clear and simple to understand, it is worth reiterating that users should avoid strictly interpreting parameters as being either an input or output. Beyond the mechanistic calculations implied by Equations (\ref{eq:t1e})-(\ref{eq:power}), all parameters are important at each analysis time, whether it's $\alpha$ to protect patients from harm, or $\beta$ to ensure that trials of effective drugs are not prematurely terminated. Therefore, an iterative process is needed to find a ``sweet spot'' that balances the various parameter values, and we encourage users to think of Fleming et al.\ and all other approaches under this framework as having different starting lines, different implications with respect to implementation, and (perhaps slightly) different focuses and tolerances for risk. 

\subsection*{Special case \#2: Rodriguez et al. (2024)}
Rodriguez et al.\ proposed to calculate the $100 \times (1-2\alpha)\%$ confidence interval (CI) for the estimated OS HR, at a time when OS data is relatively mature. Failure to exclude $\theta_0$ with the upper confidence bound is then seen as an indication of OS detriment. Their approach is equivalent to choosing a single $(\theta_0, \theta_1, \alpha, \beta)$ and solving for $(d, \theta^*)$ \citep{shan}. In other words, like traditional hypothesis testing and sample size calculation, Rodriguez et al.\ use a CI for decision making precisely at a time when a desirable set of operating characteristics can be achieved.

Rodriguez et al.\ can be generalized to multiple readouts by choosing the same set of four parameters $(\theta_0, \theta_1, \alpha, \beta)$ at each analysis time. (Users will of course have to choose at each time a different set of parameter \emph{values}, e.g., start with a larger $\alpha$ then decrease it over time.) However, in trials with very low OS event rate, it may be difficult if not impossible to reach the required number of events $d$ within a feasible time frame. In addition, timing of the first IA (or first multiple IAs) in trials of indolent cancers is often driven by an intermediate or surrogate endpoint, not OS. There may not be enough OS data at this time to thoroughly evaluate for potential harm. Even if there is, the observed number of deaths will likely differ from the prespecified number $d$, and it is unclear from this strategy what modifications will be made to address this difference. 


\subsection*{Special case \#3: Safety monitoring with standard confidence intervals}
An alternative approach to \cite{rodriguez} that one might consider is to monitor OS safety using $100 \times (1-2\alpha)\%$ CIs again but with flexibility for number of events, i.e., by choosing $(\theta_0, \theta_1, \alpha, d)$ at each analysis time. (The choice between $\theta_1$ or $\beta$ is flexible and depends on the risk one would like to quantify.)

An example of this approach is an independent data monitoring committee (IDMC) who may not have been provided with detailed instructions for OS safety monitoring and is continually using 95\% CIs for evaluation. As commented by Dr. Mikkael Sekeres at the FDA-AACR-ASA workshop, ``It is surprising how little guidance you’re often given when you’re on an IDMC... IDMC should be given good guidance on how to consider overall survival... We think that IDMC should be given specific tasks to look at overall survival at different stages of a trial and also statistical guidance on the confidence with which you can declare that a drug is either too harmful or is safe in moving forward'' \citep{aacr}. 

However, it is challenging for 95\% CIs to rule out harm early on in a trial when OS data is immature and CIs are wide. Expecting them to do so could result in early termination of an effective treatment. Other standard confidence intervals (e.g., 90\% or 80\%) may be more appropriate for safety monitoring, but specifying $\alpha$ requires careful planning. In particular, one should consider whether or not to change $\alpha$ if a different number of deaths than $d$ is observed, and if so, how.


\subsection*{Special case \#4: Discrete thresholds}
Another approach---one that we have seen study teams consider---is to choose $(\theta_0, \theta_1, d, \theta^*)$ at each analysis time, with $\theta^*$ being some discrete threshold such as 1.0, 1.1, or 1.2. There are pragmatic reasons for choosing $\theta^*$. For one, a discrete threshold is easy to communicate, understand, and follow. This could be beneficial for an IDMC that is tasked with assessing OS detriment. In addition, the OS HR point estimate is a clinically important result that can have a strong influence on the public's perception of a drug, so there could be a desire to cap it. Exceeding $\theta^*=1.0$, in particular, implies that OS detriment \emph{was observed} in the trial overall. 

However, it is unclear by choosing $(\theta_0, \theta_1, d, \theta^*)$ what risks one is willing to tolerate and how uncertainty due to small number of events is taken into consideration, if at all. Similar to approach \#3, one should consider whether or not $\theta^*$ needs to be changed if a different number of deaths than $d$ is observed. $\theta^*=1.0$ is also not appropriate if $\theta_1$ is close to 1.0, since $\beta$ will be approximately 0.50 regardless of how many OS events there are.

\subsection*{Special case \#5: FDA guidance for evaluating cardiovascular risk in type 2 diabetes (2008)}
Our final example is to choose $(\theta_1, \alpha, \beta, d)$ at IAs and $(\theta_0, \theta_1, \alpha, d)$ at the FA. This approach is based on monitoring guidelines that the FDA has used in trials of type 2 diabetes mellitus (T2D) to evaluate cardiovascular risk \citep{fda2008}, guidelines which also served to inspire approach \#1 by Fleming et al.. The key difference between approach \#5 and \#1 is that, while approach \#1 addresses the question ``With what certainty can we rule out a particular level of OS detriment?'', approach \#5 places more emphasis on type I error control at IAs and addresses the question ``What OS detriment can we rule out with a high level of certainty?''. 

It is worth noting though that both approaches \#1 and \#5 lead to the same calculated thresholds $\theta^*$, because the thresholds are uniquely determined by the parameters $\theta_1$, $\beta$, and $d$. Thus, users do not actually need to choose between the two. Rather, they can simultaneously use both approaches to understand uncertainty and to evaluate benefit-risk. This can be especially useful when a clinically relevant $\theta_0$ has been defined a priori, but few deaths have been observed to enable robust inference with respect to $\theta_0$.

\section{Case study}
\label{sec:eg}
POLARIX was a Phase 3 randomized controlled trial that compared Pola-R-CHP and R-CHOP in patients with untreated large B-cell lymphoma (LBCL). At the time of the primary analysis for PFS and first IA for OS, the trial provided statistically significant evidence of PFS benefit (PFS HR estimate 0.73; 95\% CI 0.57 to 0.95), but immature and inconclusive OS data despite all patients having a minimum of 2 years follow-up (OS HR estimate 0.94, 95\% CI 0.67 to 1.33). Subsequently, an FDA ODAC meeting was held to discuss the benefit-risk of Pola-R-CHP, taking into consideration the uncertainty around OS due to low number of deaths. For a detailed account of POLARIX and the ODAC's eventual vote in favor of Pola-R-CHP, we refer the reader to the briefing document \citep{fda}.

\cite{fleming} illustrated their OS monitoring guideline by retrospectively applying it to POLARIX (Strategy 1, Table \ref{tab:first}). Under our proposed framework, their guideline is equivalent to choosing $(d, \theta_0, \theta_1, \beta) = (89, 1.30, 0.80, 0.1)$ at IA1, $(d, \theta_0, \theta_1, \beta) = (110, 1.30, 0.80, 0.1)$ at IA2, $(d, \theta_0, \theta_1, \beta) = (131, 1.30, 0.80, 0.1)$ at IA3, and $(d, \theta_0, \theta_1, \alpha) = (178, 1.30, 0.80, 0.025)$ at FA. For the rationale behind some of these parameters, we refer the reader to \cite{fleming}. We now illustrate and compare the remaining four approaches described in Section \ref{sec:meth} by choosing similar parameter values wherever the set of input parameters overlap with Fleming et al.'s, and choosing conventional/intuitive values wherever they do not.

\begin{sidewaystable}
\caption{Various OS monitoring strategies applied to the POLARIX trial and depicted under the proposed framework. For each strategy, the four input or “chosen” parameters at a given stage are indicated in bold. Overall operating characteristics if safety thresholds are strictly adhered to are also provided. The threshold $\theta^*$ is considered ``met'' if the observed OS log hazard ratio falls below it, indicating that the experimental treatment is safe, or at least there is insufficient evidence to claim OS detriment. Strategy 1 is identical to the case study presented by Fleming et al. (2024), minus one potential interim analysis at 60 deaths for simplicity. \label{tab:first}}
\footnotesize
\begin{center}
\begin{tabularx}{\textwidth}{Xc|XXXXXX}
\hline
\RaggedRight{OS monitoring strategy} & Stage & \# of deaths ($d$) & Treatment effect under $H_0$ ($\theta_0$) & Treatment effect under $H_1$ ($\theta_1$) & \RaggedRight{HR threshold for decision making ($\theta^*$)} & \RaggedRight{Prob. of meeting threshold(s) under $H_0$ ($\alpha$)} & \RaggedRight{Prob. of meeting threshold(s) under $H_1$ ($1-\beta$)} \\ \hline
1. Fleming et al. & IA1 & \bftab{89} & \bftab{1.30} & \bftab{0.80} & 1.050 & 0.157 & \bftab{0.900} \\
(2024) & IA2 & \bftab{110} & \bftab{1.30} & \bftab{0.80} & 1.021 & 0.103 & \bftab{0.900} \\
& IA3 & \bftab{131} & \bftab{1.30} & \bftab{0.80} & 1.001 & 0.067 & \bftab{0.900} \\
& FA & \bftab{178} & \bftab{1.30} & \bftab{0.80} & 0.969 & \bftab{0.025} & 0.900 \\
& Overall & --- & --- & --- & --- & 0.017$^a$ & 0.819$^b$ \\ \hline
2. Rodriguez et & IA & 145 & \bftab{1.30} & \bftab{0.80} & 0.990 & \bftab{0.050} & \bftab{0.900} \\
al. (2024) & FA & 178 & \bftab{1.30} & \bftab{0.80} & 0.969 & \bftab{0.025} & \bftab{0.900} \\
& Overall & --- & --- & --- & --- & 0.020$^a$ & 0.869$^b$ \\ \hline
3. Standard CIs & IA1 & \bftab{89} & \bftab{1.30} & \bftab{0.80} & 1.044 & \bftab{0.150} & 0.895 \\
 & IA2 & \bftab{110} & \bftab{1.30} & \bftab{0.80} & 1.018 & \bftab{0.100} & 0.897 \\
& IA3 & \bftab{131} & \bftab{1.30} & \bftab{0.80} & 0.975 & \bftab{0.050} & 0.872 \\
& FA & \bftab{178} & \bftab{1.30} & \bftab{0.80} & 0.969 & \bftab{0.025} & 0.900 \\
& Overall & --- & --- & --- & --- & 0.016$^a$ & 0.805$^b$ \\ \hline
\end{tabularx}
\end{center}
\end{sidewaystable}

\begin{sidewaystable}
\caption*{Table 1 (continued)}
\footnotesize
\begin{center}
\begin{tabularx}{\textwidth}{Xc|XXXXXX}
\hline
\RaggedRight{OS monitoring strategy} & Stage & \# of deaths ($d$) & Treatment effect under $H_0$ ($\theta_0$) & Treatment effect under $H_1$ ($\theta_1$) & \RaggedRight{HR threshold for decision making ($\theta^*$)} & \RaggedRight{Prob. of meeting threshold(s) under $H_0$ ($\alpha$)} & \RaggedRight{Prob. of meeting threshold(s) under $H_1$ ($1-\beta$)} \\ \hline
4. Discrete & IA1 & \bftab{89} & \bftab{1.30} & \bftab{0.80} & \bftab{1.100} & 0.215 & 0.933 \\
thresholds & IA2 & \bftab{110} & \bftab{1.30} & \bftab{0.80} & \bftab{1.050} & 0.131 & 0.923 \\
& IA3 & \bftab{131} & \bftab{1.30} & \bftab{0.80} & \bftab{1.000} & 0.067 & 0.899 \\
& FA & \bftab{178} & \bftab{1.30} & \bftab{0.80} & \bftab{1.000} & 0.040 & 0.932 \\
& Overall & --- & --- & --- & --- & 0.026$^a$ & 0.854$^b$ \\ \hline
5. FDA & IA1 & \bftab{89} & 1.59 & \bftab{0.80} & 1.050 & \bftab{0.025} & \bftab{0.900} \\
guidance in & IA2 & \bftab{110} & 1.48 & \bftab{0.80} & 1.021 & \bftab{0.025} & \bftab{0.900} \\
T2D (2008) & IA3 & \bftab{131} & 1.41 & \bftab{0.80} & 1.001 & \bftab{0.025} & \bftab{0.900} \\
& FA & \bftab{178} & \bftab{1.30} & \bftab{0.80} & 0.969 & \bftab{0.025} & 0.900 \\
& Overall & --- & --- & --- & --- & 0.017$^c$ & 0.819$^b$ \\ \hline
\multicolumn{8}{l}{$^a$Probability of meeting all thresholds (interim and final) under $H_0$.} \\
\multicolumn{8}{l}{$^b$Probability of meeting all thresholds (interim and final) under $H_1$.} \\
\multicolumn{8}{l}{$^c$Probability of meeting all thresholds (interim and final) under HR = 1.30, the HR under $H_0$ at FA.} \\
\end{tabularx}
\end{center}
\end{sidewaystable}

If there is a desire to control both type I and type II error rates, then one approach is to follow \cite{rodriguez} by choosing $(\theta_0, \theta_1, \alpha, \beta)$ over time. Strategy 2 in Table \ref{tab:first} specifies standard type I error rates that decrease over time (0.05 and 0.025) and type II error rate 0.10. The resulting thresholds $\theta^*$ are similar to those of Strategy 1 at IA3 and FA. However, the actual timing of IA in POLARIX was driven by PFS, not OS; the trial targeted 228 PFS events, eventually observing 241 PFS events along with 110 OS events. Should POLARIX have assessed for potential OS detriment then, and if so, how? Or should it have waited until 145 and 178 OS events before assessing for potential OS detriment? The latter may seem odd considering that in reality POLARIX formally assessed OS for efficacy at IA under the group sequential framework.



Strategy 3 gives back flexibility on the time of safety monitoring by using 70\%, 80\%, and 90\% CIs to rule out harm early on in POLARIX. It is equivalent to choosing $(d=89, \theta_0, \theta_1, \alpha = 0.15)$, $(d=110, \theta_0, \theta_1, \alpha = 0.1)$, $(d=131, \theta_0, \theta_1, \alpha = 0.05)$, and $(d=178, \theta_0, \theta_1, \alpha = 0.025)$. The results are more relaxed thresholds and higher probabilities of claiming OS safety when Pola-R-CHP is effective, similar to those of Strategy 1.

If 95\% CIs were to be used instead from IA1 all the way through to FA (not shown in Table \ref{tab:first}), then Strategy 3 would result in very stringent thresholds (0.858 with 89 deaths, 895 with 110 deaths, and 0.923 with 131 deaths) and substantially lower probabilities of claiming OS safety when Pola-R-CHP is in fact effective (0.629, 0.721, and 0.793). It is for this reason that we do not recommend stringent control of type I error early on in a trial, since immature OS data and wide 95\% CIs can cause unnecessary alarm and result in the wrongful termination of effective treatments.

Strategy 4 considers the use of $\theta^*=$ 1.1, 1.05, 1.0, and 1.0 as the HR thresholds for decision making over time. We chose $\theta^*$ of 1.1 for IA1 and 1.05 for IA2 based on our observation that study teams have a tendency to be conservative early on in a trial when there is little data. Specific to this case study, Strategy 4 can be seen to have similar operating characteristics as Strategy 1. However, the slightly larger type I error rates of 0.215 and 0.040 at IA1 and FA might be a safety concern. More generally, this strategy does not support a transparent discussion on the various risks because it focuses on the point estimate, which can be problematic from a statistical perspective when there are few deaths.

Strategy 5 calculates what $\theta_0$ can be ruled out at the IAs with $\alpha=0.025$ and $\beta=0.1$. We see in this case that $\theta_0=1.59$, 1.48, and 1.41 given 89, 110, and 131 deaths, respectively. As noted in Section \ref{sec:meth}, Strategies 1 and 5 result in the same calculated thresholds $\theta^*$. Therefore, a user can use both strategies at an IA to understand with what level of certainty they can rule out $\theta_0=1.30$ ($\alpha=0.157$ at IA1, $\alpha=0.1031$ at IA2, and $\alpha=0.067$ at IA3) and what OS detriment they can rule out with high certainty $\alpha=0.025$ ($\theta_0=1.59$ at IA1, $\theta_0=1.48$ at IA2, and $\theta_0=1.41$ at IA3). Using Strategy 5 alone though may be problematic, as it doesn't allow a transparent discussion on the risks associated with a clinically relevant target, and may in fact claim a low risk but for a clinically irrelevant question.


Besides the \emph{marginal} risks involved at each stage in a trial, stakeholders should also consider the \emph{overall} risk of the trial should the decision making thresholds $\theta^*$ be strictly followed. Table \ref{tab:first} calculates for each strategy the probability of consistently being deemed safe (i.e., probability of meeting all thresholds $\theta^*$) under $H_0$ and under $H_1$. It can be seen that all strategies have a high probability $(>97\%)$ of being flagged as unsafe at least once (i.e., failing to meet a threshold at least once) during trial conduct if the drug is harmful. They also have a $>$80\% probability of consistently being deemed safe if the drug is effective.

For Strategies 1, 3, 4, and 5, the family wise error rates (i.e., probability of being deemed safe at least once under $H_0$) are between 0.17 to 0.24 (not shown in Table \ref{tab:first}). If there is a desire to control these rates at a lower level, say 0.05 or 0.025, then one can expect a less than 80\% probability of consistently being deemed safe under $H_1$---a level of risk that sponsors might be unwilling to take. In the context of safety monitoring where interest is in flagging a treatment that causes OS detriment, we think the probability of consistently being deemed safe under $H_0$---or more precisely, its complement, the probability of being flagged as unsafe at least once if the drug is harmful---is a more meaningful quantity than the FWER.

\section{Discussion}
\label{sec:discussion}

With advancements in anticancer therapy, it has become increasingly difficult for clinical trials in certain disease and clinical contexts to power for OS as an efficacy endpoint. Nevertheless, OS is an important endpoint for assessing safety, and rigorous benefit-risk assessment of novel therapies requires improved prospective planning, collection, and assessment of OS data. In this paper, we proposed a framework for the pre-specified analysis of OS as a safety endpoint (Section \ref{sec:meth}). Our framework solicits users to choose, at each analysis time, 4 out of 6 parameters from classical equations for statistical hypothesis testing. In doing so, users communicate clearly their focus, priority, and tolerance for risk over the course of a trial.

We described five special cases under the proposed framework: \cite{fleming}, \cite{rodriguez}, an approach that uses standard confidence intervals to continually rule out harm, another approach that uses discrete thresholds for decision making, and a fifth approach that is based off of previous FDA monitoring guidelines for evaluating cardiovascular risk in T2D. We illustrated use of the five approaches by applying them to the POLARIX trial (Section \ref{sec:eg}). While the main objective of our paper is to increase understanding of each approach, thereby facilitating planning by sponsors of future clinical trials and improving communication between stakeholders, we leave here with some final comments and recommendations regarding their use.

Of the five approaches mentioned in this paper, we think Fleming et al.'s is the most appropriate for general use due to its balance in priority over time---which shifts from a focus on ensuring that potentially effective treatments will continue to be studied when there are few OS events, to a focus on ruling out OS detriment when more OS events have accumulated. This approach also has the advantage of making clear how decision making thresholds can be calculated at any time given the observed number of OS events.

The fifth approach based on FDA's guidance in T2D can be used in tandem with Fleming et al.\ to better evaluate benefit-risk. Together, when deaths are still slowly accruing, the two approaches address the questions, ``What OS detriment can we rule out with a high level of certainty?'' and ``With what certainty can we rule out a particular level of OS detriment?''.

Rodriguez et al.\ can be used to identify the goal of a clinical trial, the point at which OS data can be considered sufficiently mature to facilitate evaluation of OS as a safety endpoint. However, it provides no guidance on how to evaluate OS at other time points, including potential interim analyses driven by the primary endpoint. Likewise, 95\% CIs provide little utility for ruling out harm early on in a trial.

The other limitation of Rodriguez et al.---the limitation that planned vs.\ observed number of deaths are typically different---can be addressed by extending the authors' approach within our proposed framework. At the design stage, sponsors can use Rodriguez et al.\ to define and predict the number of PFS and OS events at an IA, seeking a balance between properly powering PFS and observing sufficiently mature OS. Subsequently, a threshold $\theta^*$ can be calculated based on the designed $d$. At the IA, if the observed number of deaths $d'$ is different from the designed $d$, then one can use $(\theta_0, \theta_1, \alpha, d')$ or $(\theta_0, \theta_1, \beta, d')$ to update $\theta^*$ depending on what risk one could like to quantify. This is similar to the concept of group sequential designs, where stopping boundaries are updated prior to formal readouts based on the observed numbers of deaths.

\begin{figure}
\begin{center}
\includegraphics[width=0.4\textwidth]{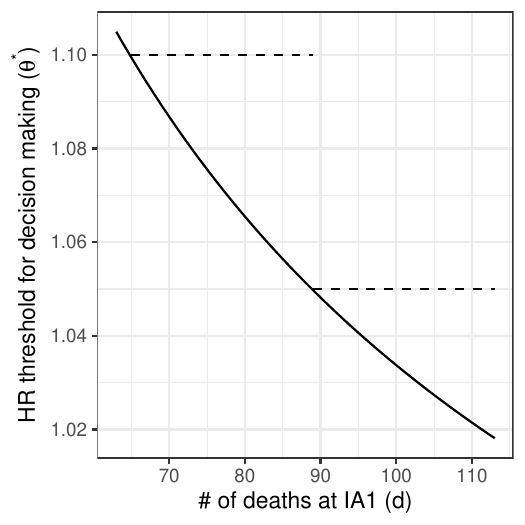}
\end{center}
\caption{\cite{fleming} and an approximately equivalent discrete threshold strategy applied to POLARIX at the first planned interim analysis with 89 expected OS events. Solid curve is the hazard ratio (HR) threshold for decision making based on \cite{fleming} with $\theta_0 = 1.3$, $\theta_1 = 0.80$, $1-\beta=0.9$, and varying potential number of observed deaths $d$. Dashed curve indicates discrete thresholds that approximate \cite{fleming} by maintaining $1-\beta \geq 0.9$ while tolerating higher levels of $\alpha$.
\label{fig:first}}
\end{figure}

More relaxed CIs (90\%, 80\%, etc.) or discrete thresholds may be appropriate provided that the risks involved are understood and acceptable. In fact, one can apply Fleming et al.\ as a first approximation, then find an approach using standard CIs or discrete thresholds that has similar operating characteristics. Such an approach could effectively balance priorities over time \emph{and} be easy to implement during a trial. Sponsors taking this approach would do well to pre-specify when and how to adjust the $\alpha$-level or discrete threshold in case the number of events deviates significantly from expectations during trial conduct (see Figure \ref{fig:first} for example).

Regardless of the approach taken, the low mortality rate in indolent diseases makes rigorous OS monitoring challenging. 
We agree with \cite{fleming} that any set of proposed thresholds should be seen as non-binding ``informative guidelines'' as opposed to black-and-white ``rules'' for continuing or stopping a trial. At a minimum, the need to monitor OS for safety should encourage stakeholders to contemplate the level of OS data that is likely to be accessible during the course of a clinical trial. Meanwhile, the additional analyses suggested by \cite{rodriguez} may be helpful for gaining a better understanding of immature OS data: analysis of biologically plausible subgroups, simulation and supplementary analyses to investigate the robustness of results under different assumptions, evaluation of other endpoints, results from other trials, use of real-world data, etc.. IDMCs may also find measures of stability or graphical summaries such as the predicted interval plot useful for understanding potential effect size estimates and associated precision with trial continuation \citep{Betensky2015, Li2009}.

Several questions remain to be addressed for the practical implementation of the proposed framework:
\begin{itemize}
    \item How should $\theta_0$ be defined, taking into account the clinical context and risk tolerance of different stakeholders? For example, patients or clinicians might tolerate greater uncertainty in OS data if a robust clinical benefit is ascertained on an intermediate/surrogate endpoint. 
    \item How should $\theta^*$ be defined given potential confounding by intercurrent events? If a substantial number of patients switch to subsequent therapy after tumor progression, then it may be desirable to relax $\theta^*$ since subsequent therapies can lead to a smaller OS difference \citep{Korn2011}. 
    \item When should OS be evaluated for safety? Should it only be evaluated when OS is sufficiently mature (however we choose to define ``mature'')? Or should it also be evaluated at key timepoints (e.g., interim and final analyses of the primary endpoint)?
\end{itemize}  

Another question that may come up is regarding non-proportional hazards (NPH). Is the proposed framework applicable under NPH? Our opinion is yes, that at least a good part of the framework can be applied \emph{if} the overall HR is still considered a meaningful treatment effect summary; for example, analysis timing can still be based on the number of deaths $d$ and $\theta^*$ can still be used as the threshold for signaling overall OS detriment. However, if it is known a priori that a NPH effect could exist, then operating characteristics should be evaluated carefully under potential patterns of NPH in order to ensure that whatever plan is proposed for OS safety analysis balances risk appropriately over time. Furthermore, supportive analyses such as piecewise exponential and cure rate models can be prospectively planned to help interpret trial results. This is similar to how some sponsors currently handle NPH for efficacy analysis. They conduct event-driven analyses using the logrank test and cox HR, but consider potential NPH when calculating the required number of events. For example, see Checkmate-274 and Keynote-966 \citep{Bajorin2021, Kelley2023}.

This paper proposed a supportive decision making framework based on a single criteria $\theta^*$. It is worth noting that \cite{rodriguez} considered having a `grey zone', which, if a trial were to fall into, would prompt further OS follow-up. Doing so could have the effect of ruling out substantial OS detriment while limiting rejection of treatments with marginal OS benefit. While we did not consider such an extension here for simplicity of comparison and presentation, we do find this idea of practical interest worth pursuing in future discussions and research.

In conclusion, it is in everyone's interest---patients, clinicians, regulators, and sponsors alike---for safe and effective cancer therapies to be developed in a timely manner. However, the low mortality rate in some disease settings can make it difficult to rigorously evaluate benefit-risk. There is a present need for stakeholders to come together, to communicate, and to align on the expectations and priorities of clinical trials with respect to the generation and interpretation of OS data. By simplifying the presentation of recent approaches for OS safety analysis and clarifying their focus, priority, and tolerance for risk over time, we believe the ``6 choose 4'' framework proposed in this paper can help to meet those needs.

\section*{Acknowledgements}
We are grateful to Emmanuel Zuber, Lisa Hampson, Arunuva Chakravartty, and other co-authors of \cite{fleming} for their time and valuable feedback during the development of this manuscript. We would also like to thank two referees and an Associate Editor who helped to clarify and improve the presentation of this paper.

\section*{Disclosure statement}
The authors report that there are no competing interests to declare. All authors are employees of Roche and own stocks in this company.

\bibliographystyle{agsm.bst}
\bibliography{main.bib}

\end{document}